\documentclass[aip, jcp, reprint]{revtex4-1}
\usepackage{bm}
\usepackage{amsmath}
\usepackage{amsfonts}
\usepackage[pdftex]{graphicx}
\usepackage[utf8]{inputenc}

\begin{document}

\title{Bayesian statistical modelling of microcanonical melting times at the superheated regime}

\author{Sergio Davis}
\email{sdavis@cchen.cl}
\affiliation{Comisión Chilena de Energía Nuclear, Casilla 188-D, Santiago, Chile}
\affiliation{Departamento de F\'isica, Facultad de Ciencias Exactas, Universidad Andres Bello. Sazi\'e 2212, piso 7, 8370136, Santiago, Chile.}

\author{Claudia Loyola}
\affiliation{Departamento de F\'isica, Facultad de Ciencias Exactas, Universidad Andres Bello. Sazi\'e 2212, piso 7, 8370136, Santiago, Chile.}

\author{Joaquín Peralta}
\affiliation{Departamento de F\'isica, Facultad de Ciencias Exactas, Universidad Andres Bello. Sazi\'e 2212, piso 7, 8370136, Santiago, Chile.}

\date{\today}

\begin{abstract}
Homogeneous melting of superheated crystals at constant energy is a dynamical process, believed to be triggered by the accumulation 
of thermal vacancies and their self-diffusion. From microcanonical simulations we know that if an ideal crystal is prepared at a 
given kinetic energy, it takes a random time $t_w$ until the melting mechanism is actually triggered. In this work we have studied in detail the 
statistics of $t_w$ for melting at different energies by performing a large number of Z-method simulations and applying state-of-the-art methods 
of Bayesian statistical inference. By focusing on the short-time tail of the distribution function, we show that $t_w$ is actually gamma-distributed rather 
than exponential (as asserted in previous work), with decreasing probability near $t_w \sim 0$. We also explicitly incorporate in our model the 
unavoidable truncation of the distribution function due to the limited total time span of a Z-method simulation. The probabilistic model presented 
in this work can provide some insight into the dynamical nature of the homogeneous melting process, as well as giving a well-defined practical procedure 
to incorporate melting times from simulation into the Z-method in order to reduce the uncertainty in the melting temperature.
\end{abstract}

\maketitle

\section{Introduction}
\label{introduction}

It is well known that, under carefully controlled conditions, a solid can be heated above its melting temperature
without triggering the melting process. This is known as superheating, and in it the solid enters a metastable state
with interesting kinetic properties~\cite{Forsblom2005}. It is also possible to achieve such an homogeneous melting by
ultrafast laser pulses~\cite{Siwick2003}. As a complement to experiments, observation and characterization of this superheated state 
is especially clear in microcanonical simulations, particularly within the framework of the Z-method~\cite{Belonoshko2006}. By using this 
methodology, it has been possible to establish the role of thermal vacancies~\cite{Belonoshko2007,Pozo2015} and their
self-diffusion~\cite{Bai2008, Davis2011, Amit2014}.

The Z-method, in fact, relies on the superheating metastable state to determine the melting temperature $T_m$. It 
establishes that in microcanonical isochoric simulations there is a maximum energy $E_{LS}$ ($LS$ stands for Limit 
of Superheating), related to a temperature $T_{LS}$, that can be given to the crystal before it spontaneously melts. 
Beyond this point the solid unavoidably melts, showing a sudden, sharp drop in temperature, corresponding 
to the kinetic energy consumed as latent heat of melting. When a system melts at an energy $E_{LS}+\Delta E$ (with
$\Delta E \rightarrow 0$), the final temperature $T$ approaches $T_m$. Despite the current lack of a complete
theory of first-order phase transitions and metastable states, there is plenty of accumulated evidence of the accuracy
of the Z-method~\cite{GonzalezCataldo2016, Balashenko_2013, Belonoshko_2009}. However, some issues remain concerning the 
uncertainties associated to $T_m$ and $T_{LS}$ due to the use of short simulation times. One important source of
uncertainty is the random distribution of the elapsed time $t_w$ from the beginning of the simulation until the melting 
process is triggered, even in a microcanonical setting. In previous work, Alfè \emph{et al}~\cite{Alfe2011} reported 
statistics of $t_w$ obtained from molecular dynamics (MD) simulations and postulated that $t_w$ is exponentially distributed, 
and therefore the most probable value of $t_w$ is close to zero for all temperatures.

In this work, we performed a large number of microcanonical, isochoric simulations in order to generate statistically
independent samples of $t_w$ in a wide range of initial temperatures. We focused particularly in shorter times and found
that these times are not exponentially distributed but gamma-distributed. Accordingly, we present a precise model that 
reproduces the gamma shape and scale parameters as functions of the initial temperature.

The rest of the paper is organized as follows. In section \ref{computational} we give a detailed description of the simulations 
that were used to determine the melting temperature and the samples of waiting times ($t_w$). Section \ref{results}
describes the probability models proposed to represent the statistical distribution of $t_w$ and assesses their
likelihood given our simulated data. Next, sections \ref{modelling} and \ref{trunc_gamma} present a concrete model for the 
distribution of $t_w$ as a function of temperature fitted to our simulation data. Finally, section \ref{concluding} discusses 
the implications of our results for the process of melting in the superheating state for homogeneous solids.

\section{Computational Procedure}
\label{computational}

As a simple model of solid we considered a high-density, face-centered cubic (FCC) argon crystal, having 500 atoms and
lattice constant $a$=4.2\AA. This ideal structure was the starting point for all simulations. The use of such a high
density (and therefore pressure) increases the superheating effect, revealing the melting process in a more salient way. 
All simulations were performed using the LPMD~\cite{Davis2010} molecular dynamics package. For every case we use a total 
simulation time $\tau_0$=50 ps and a timestep $\Delta t=$0.5 fs with a Beeman integrator for Newton's equations of motion. 
The interatomic potential used in the simulations corresponds to a standard Lennard-Jones model with $\epsilon/k_B=119.8$K 
and $\sigma=3.41$\AA, as used in previous works on the Z-method~\cite{Belonoshko2006,Belonoshko2007}.

\begin{figure}[h!]
\begin{center}
\includegraphics[width=0.95\columnwidth]{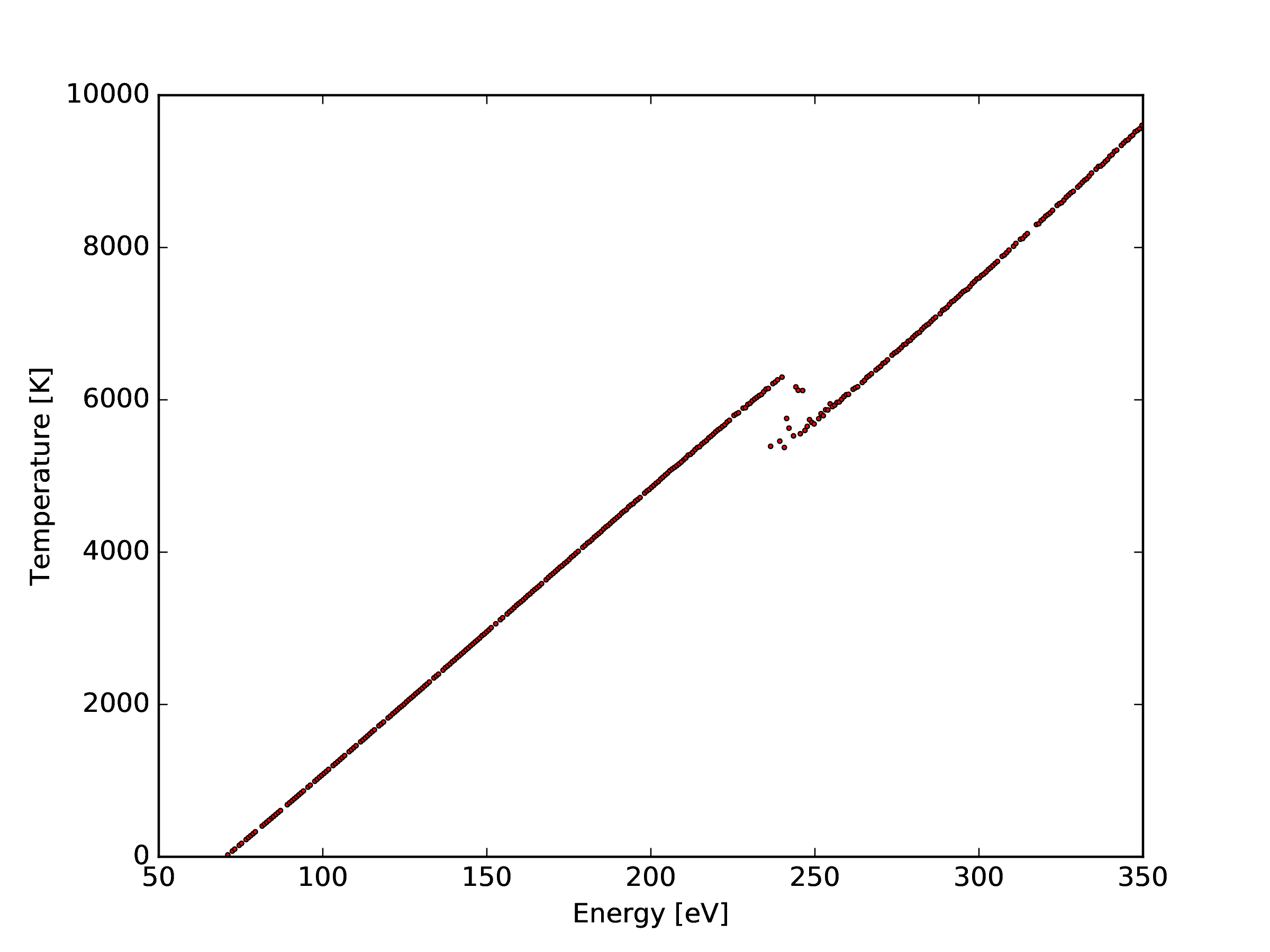}
\caption{Isochoric (Z) curve obtained from 400 different 50 ps simulations of high-density Ar, with initial 
temperatures ranging from 50 to 20000 K. From this Z-curve we can determine $T_m \sim$ 5376 K and $T_{LS} \sim$ 6291 K. 
A clear drop of temperature is observed near the critical energy $E_{LS}\sim$ 240 eV.}
\label{fig-Zmeth}
\end{center}
\end{figure}

A set of 400 simulations with temperatures ranging from 50 K to 20000 K have been used to draw the isochoric (Z) curve
for this high-density argon model, in a standard application of the Z-method. These results are shown in Fig. \ref{fig-Zmeth}.

From the isochoric, Z-shaped curve, we have estimated the melting temperature $T_m$ and the superheating limit
$T_{LS}$. The sharp inflection at the higher temperature (6291 K) corresponds to $T_{LS}$ and the lower
inflection (5376 K) corresponds to $T_m$. We can also determine $E_{LS}$ at approximately 240 eV.

Once $T_{LS}$ is estimated, at least 5000 classical MD runs were performed for 24 different initial temperatures $T_0$ in an interval 
from $T_0=$12075 K to $T_0=$13300 K, in order to collect statistically independent samples of waiting times. As it is known, by 
the equipartition theorem the initial temperature drops to around half the initial value when no velocity rescaling procedure or 
thermostat is used. If the specific heat at constant volume $C_v$ is independent of $T$ for the solid branch of the
isochore, the temperature $T(E)$ is a linear function of the total energy $E$. Because $E$ is given by 

\begin{equation}
E = \frac{3N}{2}k_B T_0 + \Phi_0 = E_0 + C_v T,
\end{equation}
we see that $T = m\;T_0 + b$, in our case, $m$=0.40535 $\pm$ 0.004633 and $b$=1324.034 $\pm$ 56.37 K. The specific heat
per atom is $3/(2m)$ in units of $k_B$, about 3.7.

\begin{figure}[h!]
\begin{center}
\includegraphics[width=0.95\columnwidth]{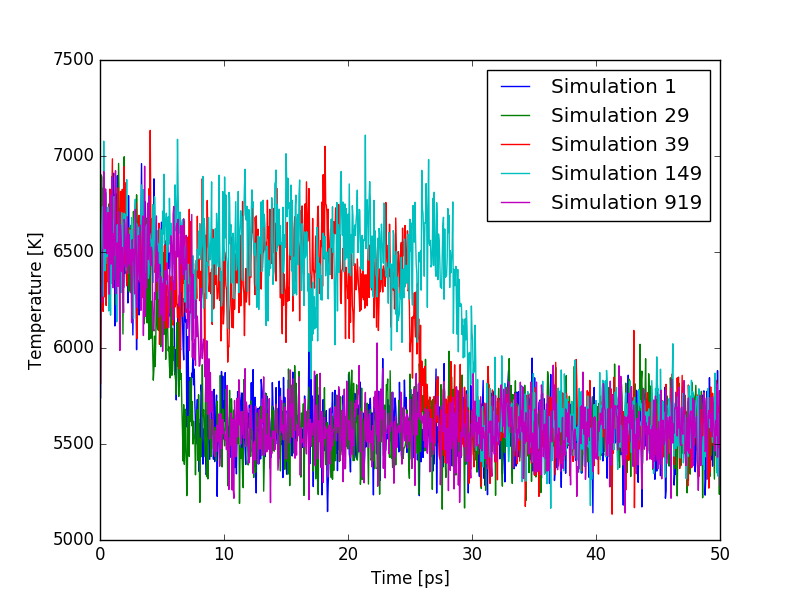}
\caption{Instantaneous temperature $T(t)$ from five different simulations at an initial temperature $T_0=$12800 K. 
The melting process is triggered at different times $t_w$ between 5 and 30 ps for this particular energy.}
\label{fig-melt}
\end{center}
\end{figure}

As an example of the dynamical behavior of the instantaneous temperature $T(t)$ at the moment of melting, four realizations of
MD simulation are shown, superimposed, in Fig. \ref{fig-melt} for an initial temperature $T_0$=12800 K. Here we see the large
variability of $t_w$ in those samples, ranging from approx. 5 ps to 30 ps.

From the MD simulations with duration $\tau_0$=50 ps we only considered for the collected statistics the values of $t_w$
such that $t_w \leq \tau_0$ (i.e. the simulatione were melting is not observed in the time window $\tau_0$ were ignored). Because of this, we are in
fact inferring the truncated distribution $P(t_w | t_w \leq \tau_0, T)$. For all simulations were melting is observed we extracted the temperature 
before and after the melting process, denoted by $T_{\text{solid}}$ and $T_{\text{liquid}}$ respectively, and the
melting time $t_w$. This was done via a simple automated least-squares fitting using the following ``discontinuous step'' model, 

\begin{equation}
   T(t) = 
   \begin{cases}
    T_{\text{solid}} & \;\text{if}\; t \leq t_w , \\
    T_{\text{liquid}} & \;\;\text{otherwise}.
   \end{cases}
\end{equation}

For the lowest part of the range of initial temperatures, more precisely between 12000 K and 12250 K, 10000 simulations
per temperature were performed instead of 5000. This is of course because the melting events are much less probable once we approach 
$T_{\text{solid}}=T_{LS}$ from above. In this way we have collected at least 200 samples of waiting times for each $T_0$.

In the next section we present a statistical description of $t_w$ for different initial temperatures, and a
comparison of the exponential, gamma, log-normal and kernel density estimation (KDE) models against our data.

\section{Results}
\label{results}

\begin{figure}[h!]
\begin{center}
\includegraphics[width=0.95\columnwidth]{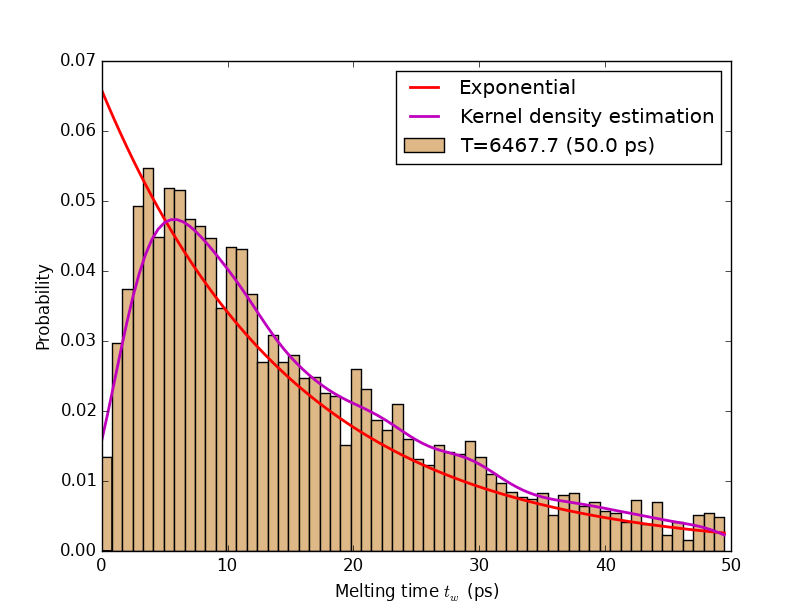}
\caption{Probability density function for waiting times at $T_0$=12690 K ($T$=6467.7 K).}
\label{fig_dist1}
\end{center}
\end{figure}

\begin{figure}[h!]
\begin{center}
\includegraphics[width=0.95\columnwidth]{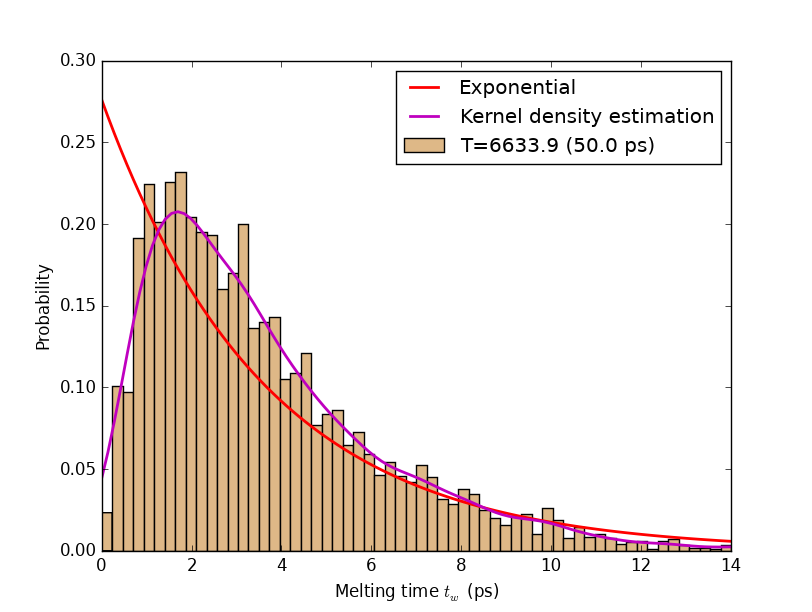}
\caption{Probability density function for waiting times at $T_0$=13100 K ($T$=6633.9 K).}
\label{fig_dist2}
\end{center}
\end{figure}

\begin{figure}[h!]
\begin{center}
\includegraphics[width=0.95\columnwidth]{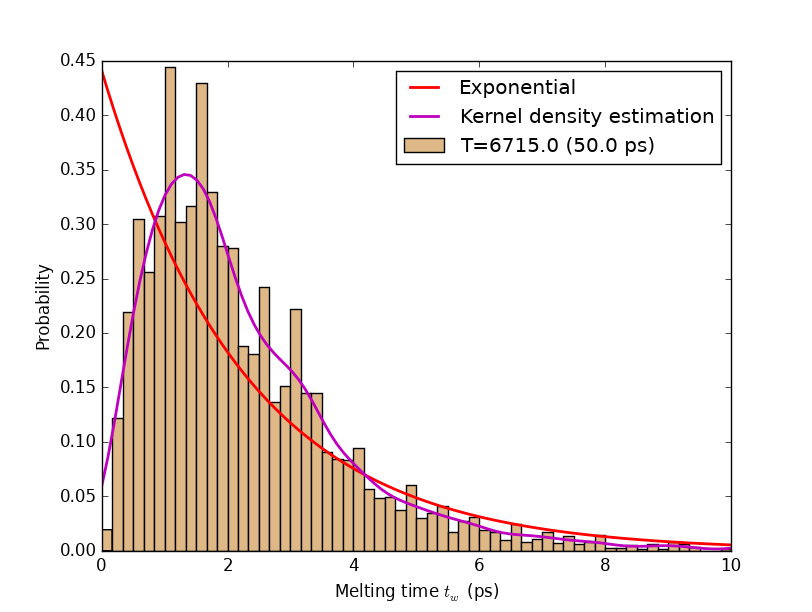}
\caption{Probability density function for waiting times at $T_0$=13300 K ($T$=6715.0 K).}
\label{fig_dist3}
\end{center}
\end{figure}

We have performed, for all 24 different initial temperatures $T_0$, a statistical analysis of the samples of melting
time $t_w$ collected from several thousand MD simulations as described in the previous section. Figs.~\ref{fig_dist1},
\ref{fig_dist2} and \ref{fig_dist3} show the histograms of $t_w$ samples for increasing $T_{\text{solid}}$ above
$T_{LS}$, together with their respective maximum likelihood exponential fit, with the exponential distribution given by 

\begin{equation}
P(t_w | \lambda ) = \lambda \exp(-\lambda t_w)
\end{equation}
with $\lambda$ a scale parameter, and a kernel density estimation.

We see in all cases that the exponential model cannot reproduce the left tail of the histogram, overestimating the
probability of $t_w=0$. In contrast, this behavior is correctly reproduced by the kernel density method. Our results show 
that there is no instantaneous melting process, but a certain latency exists in which presumably some particular
conditions in the crystal have to be set up. This can be explained by the fact that the collapse of the solid must 
involve the crossing of an energy barrier via a random walk in phase space.

As expected, the most probable value of $t_w$ (i.e. its mode) decreases with the initial temperature, and the gap between the
exponential and KDE models widens. We search, therefore, for an alternative statistical model with the same asymmetry
observed in the histograms (long right tail) and with non-zero mode. We propose the gamma and log-normal models as
candidates, which are assessed against the data in the next section.

\section{Statistical models of waiting times and model comparison}
\label{modelling}

At this point we consider as candidate statistical models for the waiting time $t_w$ the gamma distribution, given by

\begin{equation}
P(t_w | k, \theta) = \frac{\exp(-t_w/\theta)t_w^{k-1}}{\Gamma(k)\theta^k},
\label{eq_gamma}
\end{equation}
where $k$ and $\theta$ are the shape and scale parameters, respectively, and the log-normal distribution, given by 

\begin{equation}
P(t_w | \mu, \sigma) = \frac{1}{\sqrt{2\pi}\sigma t_w}\exp\left(-\frac{1}{2\sigma^2}(\ln t_w - \mu)^2\right).
\end{equation}

\begin{figure}[h!]
\begin{center}
\includegraphics[width=0.95\columnwidth]{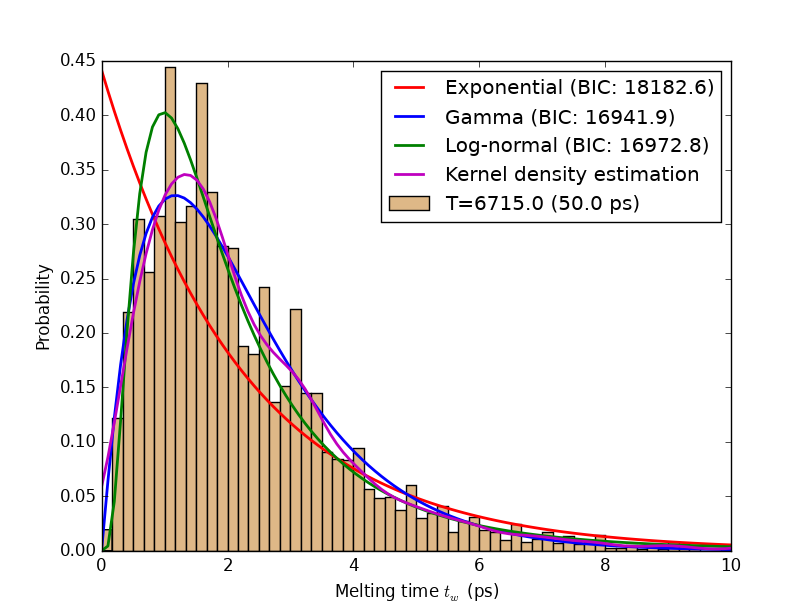}
\caption{Comparison of exponential, gamma and log-normal models for waiting times at $T$=6715 K. The Bayesian
Information Criterion (BIC) is reported for all models. Also a Gaussian kernel density estimation is included for
reference.}
\label{fig-bic}
\end{center}
\end{figure}

The comparison between these two models plus the exponential model, together with the kernel density estimation, is shown in Fig. \ref{fig-bic}. 
In this figure the goodness-of-fit of the models is also evaluated quantitatively using the Bayesian Information Criterion (BIC)~\cite{Schwarz1978}. 
This criterion is based on computing the quantity

\begin{equation}
\text{BIC} = -2\ln L(\bm{\lambda_0}) + n_p\ln n_d,
\end{equation}
where $L(\bm{\lambda})=P(\text{data}|\bm{\lambda})$ is the likelihood function for the model, $\bm{\lambda_0}$ is the
parameter vector for which $L$ is maximum, $n_p$ is the number of parameters in the model and $n_d$ is the number of
data points. Under this definition, the model with lowest BIC should be the most appropriate to represent the data. The
first term favors models with high likelihood, while the second term penalizes the models with large number of
parameters in order to avoid overfitting.

We see that the lowest value of BIC is obtained by the gamma model, which is also visually the closest to the KDE
reference model. Accordingly, we have chosen the gamma model as the basis of our further statistical description in the
next section, where we incorporated the effect of truncation due to the limited duration of the MD simulations in the Z-method.

\section{The truncated gamma model}
\label{trunc_gamma}

As was described previously, in every Z-method simulation there is an intrinsic truncation of the waiting times due to
the finite duration of each run. We can incorporate this effect of truncation in the gamma model given by Eq.
\ref{eq_gamma} as follows. The truncated probability density of $t_w$ is the conditional probability $P(t_w|t_w \leq \tau_0, T)$, 
which due to Bayes' theorem can be written as

\begin{equation}
P(t_w|t_w \leq \tau_0, T) = \frac{P(t_w|T)P(t_w \leq \tau_0|t_w, T)}{P(t_w \leq \tau_0|T)},
\label{eq_ptrunc}
\end{equation}
where $P(t_w|T)$ is given by Eq. \ref{eq_gamma} with $k=k(T)$ and $\theta=\theta(T)$. Moreover, the probability $P(t_w
\leq \tau_0|t_w, T)$ does not depend on $T$ and takes the value one if $t_w \leq \tau_0$, zero otherwise. So, it can be written 
as $\Theta(\tau_0 - t_w)$ for all values of $T$. Imposing normalization of the left-hand side of Eq. \ref{eq_ptrunc},

\begin{align}
P(t_w \leq \tau_0|T) &= \int_0^\infty dt_w P(t_w|T)\Theta(\tau_0-t_w) \nonumber \\
                     &= \int_0^{\tau_0} dt_w P(t_w|T).
\label{eq_norm}
\end{align}

\noindent
Combining Eqs. \ref{eq_gamma}, \ref{eq_ptrunc} and \ref{eq_norm} finally we arrive at

\begin{equation}
P(t_w |t_w < \tau_0, T) = \frac{\exp(-t_w/\theta(T))t_w^{k(T)-1}\Theta(\tau_0-t_w)}{\Gamma_\text{inc}(k(T); 0 \rightarrow \frac{\tau_0}{\theta(T)})\theta(T)^{k(T)}}
\end{equation}
where $\Gamma_\text{inc}(k; a \rightarrow b)$ is the incomplete gamma function, given by

\begin{equation}
\Gamma_\text{inc}(k; a \rightarrow b) = \int_a^b dt \exp(-t)t^{k-1}.
\end{equation}

Using this truncated model, we can compute several statistical properties of the waiting times. First, the probability
of observing melting before a time $\tau_0$ at temperature $T$ is given by

\begin{equation}
P(t_w < \tau_0 | T) = \frac{\Gamma_\text{inc}(k(T); 0 \rightarrow \frac{\tau_0}{\theta(T)})}{\Gamma(k(T))}.
\label{eq_freq_melting}
\end{equation}

Similarly, the most probable waiting time ${t_w}^*$ at a given $T$ is given by ${t_w}^*=(k(T)-1)\theta(T)$, while the
average value of $t_w$ given a simulation time $\tau_0$ at a temperature $T$ is 

\begin{equation}
\big<t_w\big>_{\tau_0,T} = \theta(T)\left[\frac{\Gamma_{\text{inc}}(k(T)+1; 0 \rightarrow
\frac{\tau_0}{\theta(T)})}{\Gamma_{\text{inc}(k(T); 0 \rightarrow \frac{\tau_0}{\theta(T)})}}\right],
\end{equation}
which goes to $k(T)\theta(T)$ as $\tau_0 \gg \theta(T)$, as expected.


All that remains for a complete statistical description is the estimation of the functions $k(T)$ and $\theta(T)$. We
address this problem by means of Bayesian inference, first using a Bézier 50-parameter model (described in Appendix
\ref{appendix_bezier}) and a 7-parameter model, which we describe below.

\newpage

\begin{figure}[h!]
\begin{center}
\includegraphics[width=1.05\columnwidth]{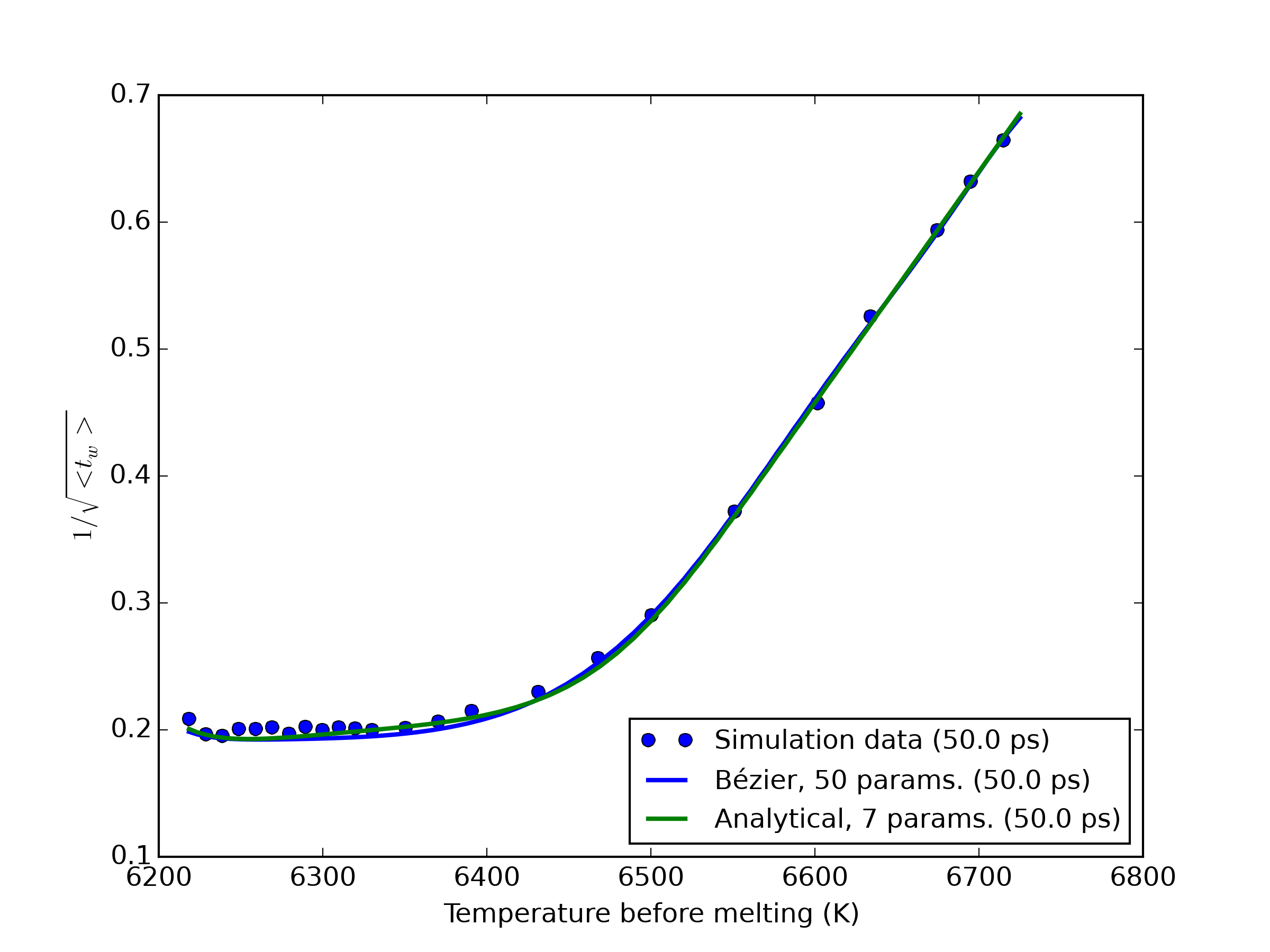}
\end{center}
\caption{The quantity $1/\sqrt{\big<t_w\big>_T}$ as a function of $T$. Circles represent simulation points with
$\tau_0$=50 ps, while the blue and green lines are the Bézier and 7-parameter analytical models proposed in this work,
respectively. We see that the linear 
behavior for high temperatures is no longer valid below a certain temperature $T^*$, in this case approximately 6550 K.}
\label{fig_alfe}
\end{figure}

Based on the original model by Alfè \emph{et al}, we propose that for sufficiently high temperatures so that $\tau_0 \gg
\theta(T)$, i.e. when truncation effects are negligible and the probability of melting approaches one, the quantity 

\begin{equation}
a(T)=\frac{1}{\sqrt{\big<t_w\big>_T}}
\label{eq_alfe}
\end{equation}
increases linearly with $T$. However, at low temperatures it saturates, as seen in the simulation data (Figure \ref{fig_alfe}, circles).

Accordingly, we will define an ``inflection temperature'' $T^*$ that separates both regimes, in such a way that $a(T)$ is given by

\begin{equation}
a(T) = 
   \begin{cases}
    C + R(T-T_{\text{base}})^\gamma & \;\text{if}\; T \leq T^* , \\
    \alpha T - \beta & \;\;\text{otherwise}.
   \end{cases}
\label{eq_model1}
\end{equation}
where the parameter $T_{\text{base}}$ is a reference temperature, fixed here at 6200 K, and the parameters $R$ and $C$
are fixed by the continuity of $a(T)$ and its derivative at $T^*$ as

\begin{eqnarray}
R = \frac{\alpha}{\gamma(T^*-T_{\text{base}})^{\gamma-1}}, \nonumber \\
C = \alpha T^* - \beta - R(T^*-T_{\text{base}})^\gamma.
\label{eq_model2}
\end{eqnarray}

As $\big<t_w\big>_T = k(T)\theta(T)$ for a gamma-distributed $t_w$, it follows that 

\begin{equation}
k(T) = \frac{1}{\theta(T) a(T)^2}.
\label{eq_model3}
\end{equation}

Now, the behavior of $\theta(T)$ is monotonically decreasing with $T$, and on first thought an Arrhenius law would be
expected. In fact, however, it deviates from an exponential and we added a quadratic dependence on $\ln \theta(T)$,

\begin{equation}
\ln \theta(T) = \ln \theta_0 -\left(\frac{T-T_{\min}}{\eta}\right) - \left(\frac{T-T_{\min}}{\zeta}\right)^2.
\label{eq_model4}
\end{equation}

\begin{table}
\begin{tabular}{|c|c|c|}
\hline
Parameter & Confidence interval & Uncertainty \% \\
\hline
$T^*$ (K)                       & 6550.02 $\pm$ 1.80 & 0.0275\% \\
\hline
$\gamma$                        & 1.8487 $\pm$ 0.0051 & 0.276\% \\
\hline
$\theta_0$ (ps)                 & 2667.51 $\pm$ 96.03 & 3.600\% \\
\hline
$\eta$ (K)                      & 40.48 $\pm$ 0.39 & 0.963\% \\
\hline
$\zeta$ (K)                     & 231.75 $\pm$ 2.23 & 0.962\% \\
\hline
$\alpha$ (ps$^{-1/2}$K$^{-1}$)  & 1.8198$\times$10$^{-3}$ $\pm$ 1.02$\times$10$^{-6}$ & 0.056\% \\
\hline
$\beta$ (ps$^{-1/2}$)           & 11.5446 $\pm$ 0.0065 & 0.0563\% \\
\hline
\end{tabular}
\caption{Estimated parameters for the analytical model given by Eqs. \ref{eq_model1} to \ref{eq_model4}. The confidence
intervals reported are constructed using the mean and standard deviation of Markov Chain Monte Carlo samples from the posterior distribution.}
\label{tbl_params}
\end{table}

Finally, we have that the functions $k(T)$ and $\theta(T)$ are described by 7 parameters in total, namely $T^*$, $\gamma$,
$\theta_0$, $\eta$, $\zeta$, $\alpha$ and $\beta$. These are in fact hyperparameters for the gamma distribution, and were estimated given 
our simulation data using a Bayesian Markov Chain Monte Carlo (MCMC) procedure, described in detail in Appendix \ref{appendix_MC}. 

\begin{figure}[h!]
\begin{center}
\includegraphics[width=1.05\columnwidth]{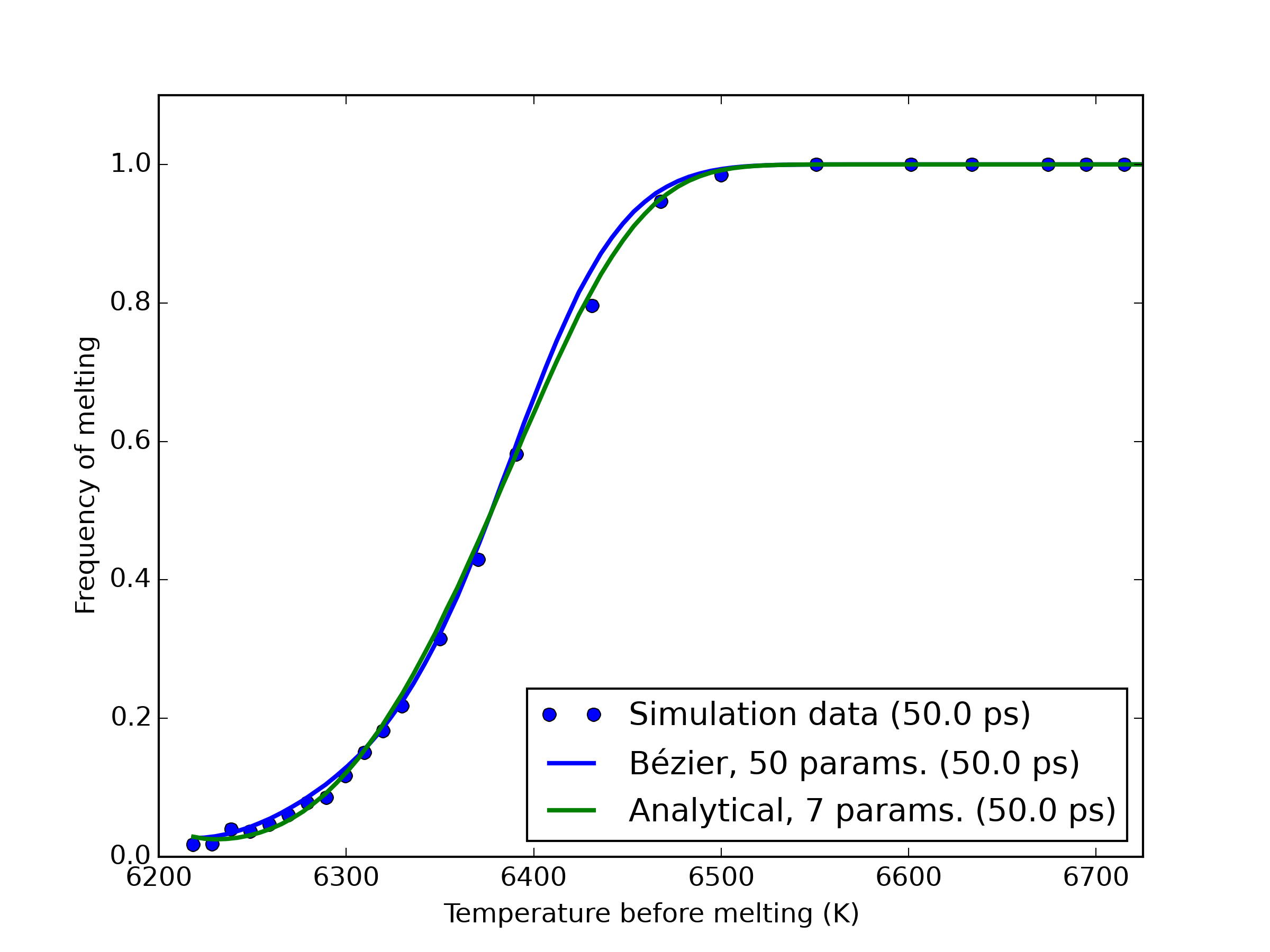}
\end{center}
\caption{Frequency of melting (Eq. \ref{eq_freq_melting}) as a function of $T$ for the truncated gamma model. Circles represent the
simulation data with $\tau=$50 ps, while the blue and green lines are the Bézier and 7-parameter analytical models proposed in this work, respectively.}
\label{fig_model_freq}
\end{figure}

\begin{figure}[h!]
\begin{center}
\includegraphics[width=1.05\columnwidth]{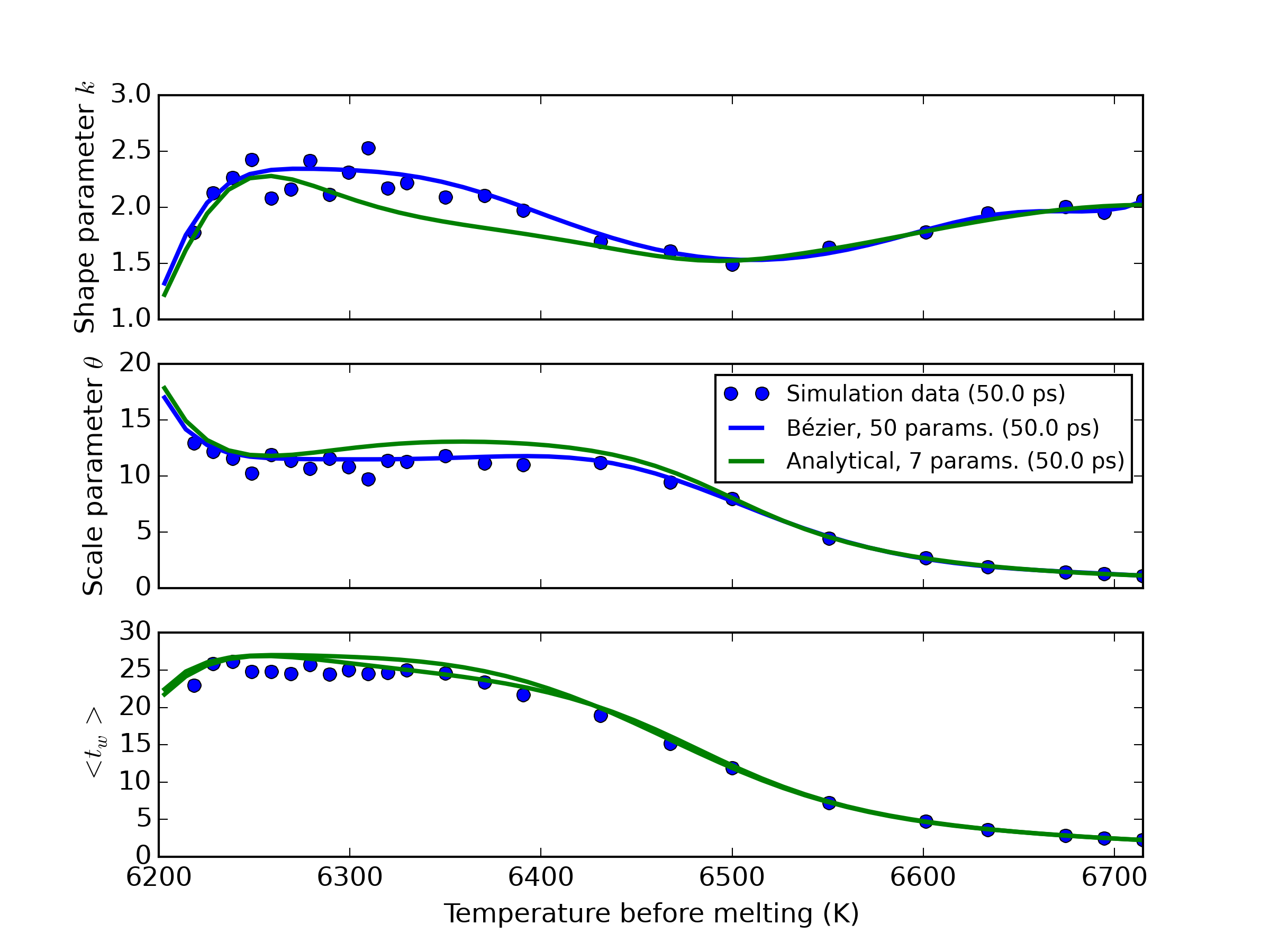}
\end{center}
\caption{Parameters $k(T)$, $\theta(T)$ and the expected waiting time $\big<t_w\big>_{T,\tau_0}$ for the truncated gamma model. Circles 
represent the simulation data with $\tau=$50 ps, while the blue and green lines are the Bézier and 7-parameter analytical models proposed in this work, respectively.}
\label{fig_model_params}
\end{figure}

The confidence intervals for all parameters from this procedure are summarized in Table \ref{tbl_params}. For all
parameters the uncertainty is below 4\%, and in some cases well below 1\%. Fig. \ref{fig_model_freq} shows 
the agreement of the same models on the frequency of melting as a function of $T$, while Fig. \ref{fig_model_params} shows the agreement
of the two models (Bézier and 7-parameter) against the simulation data on the truncated parameters $k(T)$, $\theta(T)$
and $\big<t_w\big>_{T,\tau_0}$. In both cases we see a remarkable agreement with the simulation, with the Bézier model
providing the best fit due to the large number of parameters. The 7-parameters model seems to be precise enough in its predictions for practical use. 

\begin{figure}[h!]
\begin{center}
\includegraphics[width=1.05\columnwidth]{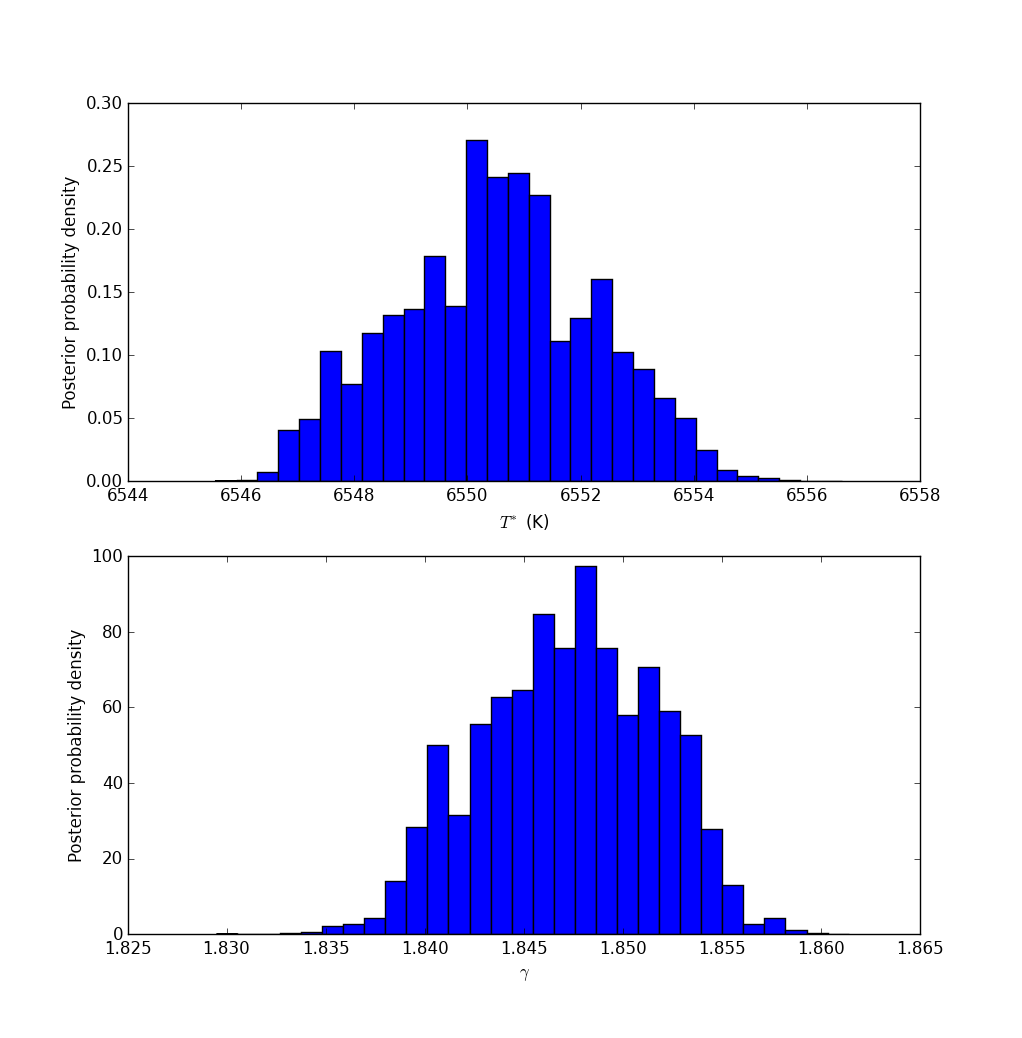}
\caption{Bayesian posterior densities for the parameters $T^*$ and $\gamma$.}
\label{fig_posterior_1}
\end{center}
\end{figure}

\begin{figure}[h!]
\begin{center}
\includegraphics[width=1.05\columnwidth]{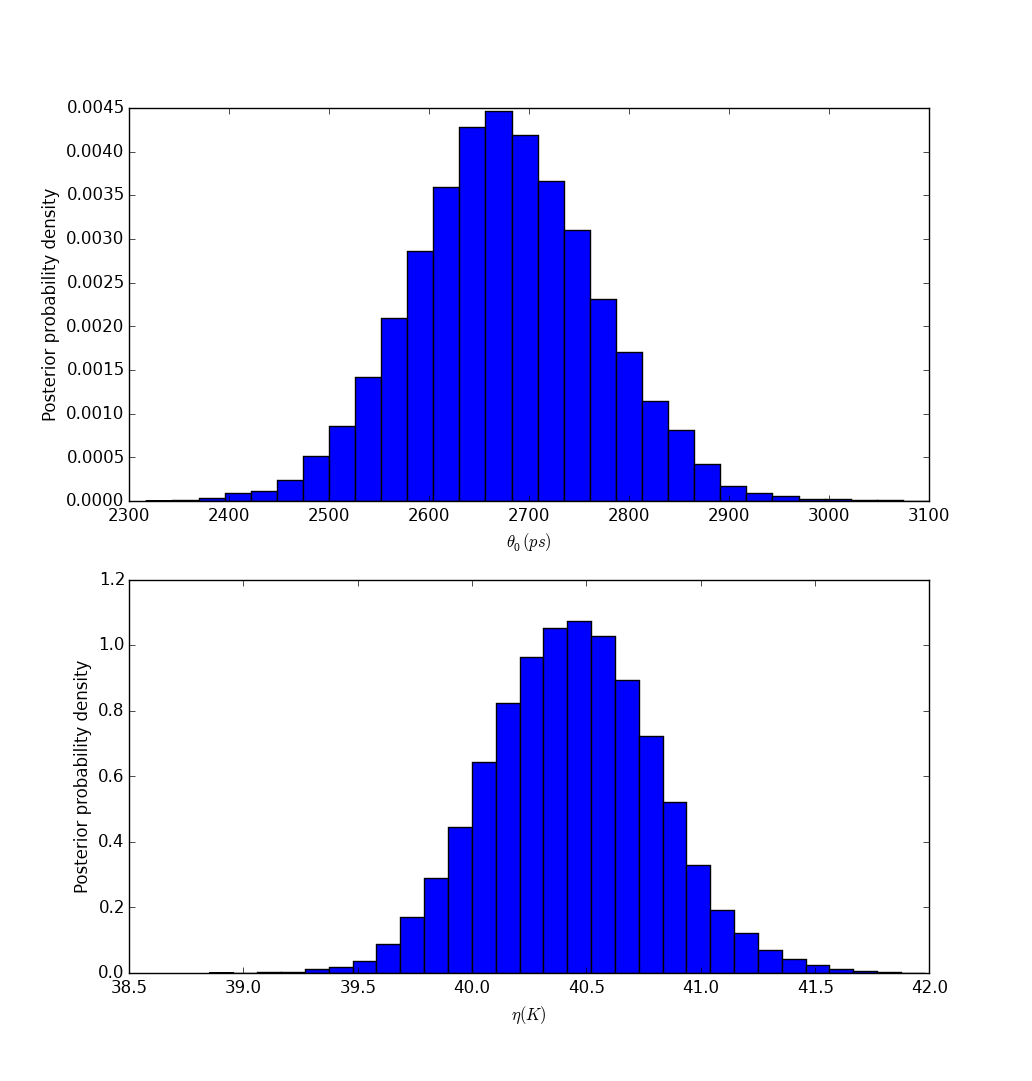}
\caption{Bayesian posterior densities for the parameters $\theta_0$ and $\eta$.}
\label{fig_posterior_2}
\end{center}
\end{figure}

\begin{figure}[h!]
\begin{center}
\includegraphics[width=1.05\columnwidth]{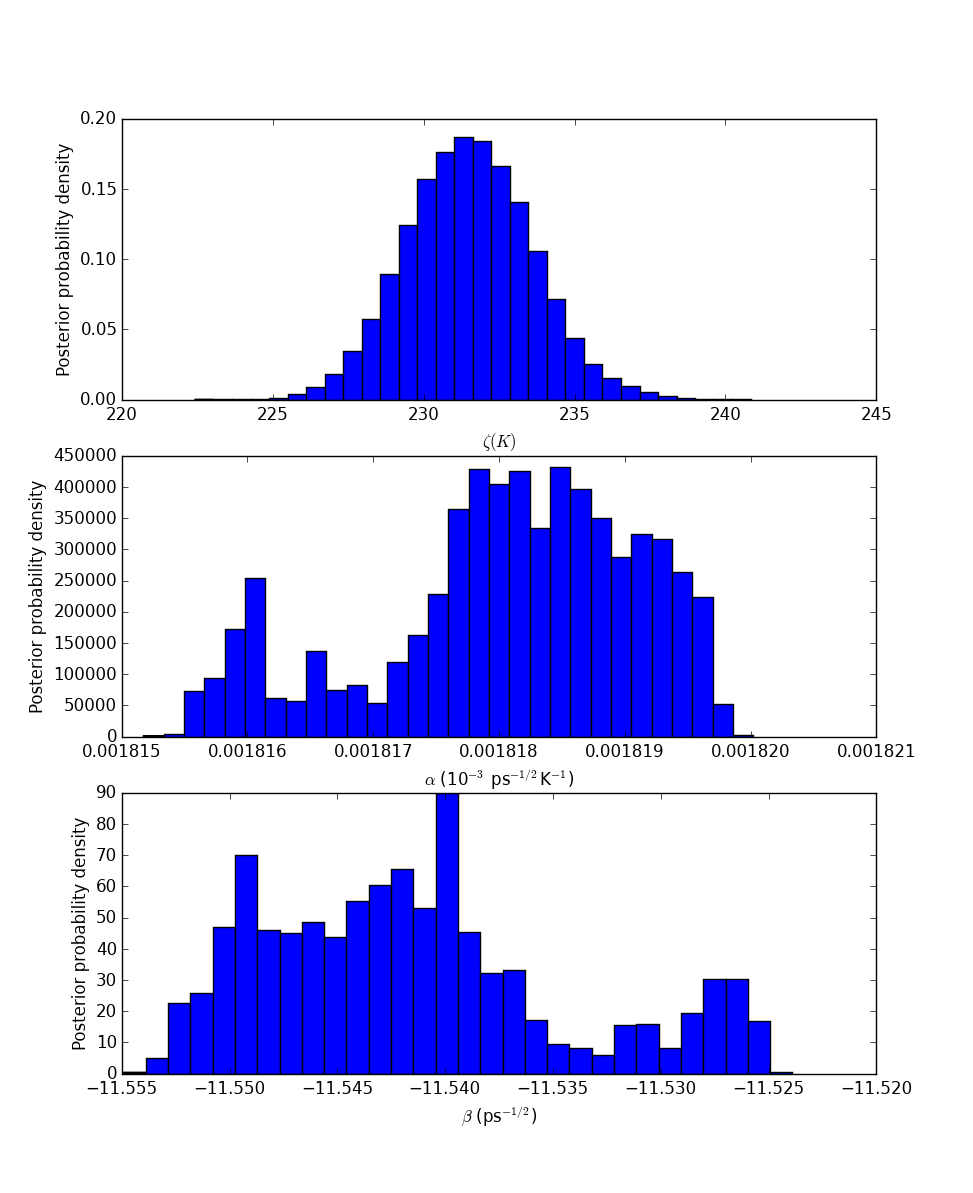}
\caption{Bayesian posterior densities for the parameters $\zeta$, $\alpha$ and $\beta$.}
\label{fig_posterior_3}
\end{center}
\end{figure}

Figs. \ref{fig_posterior_1}, \ref{fig_posterior_2} and \ref{fig_posterior_3} show the Bayesian posterior distribution
functions sampled using the MCMC procedure. For all parameters except $\alpha$ and $\beta$ the posterior distribution
has been resolved clearly, and is unimodal with a well-defined shape. In the case of $\alpha$ and $\beta$ the uncertainty 
is so small (approx. 0.056\%) that the MCMC procedure does not resolve the shape of the peak. However, this is of no
importance, because for such a small uncertainty a point estimate contains essentially all the information about the distribution.

\section{Concluding remarks}
\label{concluding}

We have presented a detailed statistical description of the waiting times in homogeneous, microcanonical melting as a
function of initial energy. The 7-parameter model, together with their uncertainties, was constructed by using a large set of 
80000 independent molecular dynamics simulations and the application of Bayesian inference techniques.
Our results shows that the waiting time $t_w$ is gamma-distributed, not exponential, with a well-defined probability maximum at 
non-zero times. We interpreted this as the absence of instantaneous melting due to the intrinsic latency caused by the
initial random exploration in phase space, prior to the crossing of an energy barrier. 

The model we present here could be of use not only for further understanding of the microcanonical melting in ideal
conditions, but also in situations where ultrafast melting occurs, in which the time and length scales are so short that
essentially a microcanonical situation arises. Two examples with technological relevance are laser-induced
melting~\cite{Plech2004,Musumeci2010,Siwick2003} and melting produced by radiation~\cite{Shirokova2013,InestrosaIzurieta2015} impacting on surfaces.

It would be interesting to explore the possibility that the parameters of the model given in Table \ref{tbl_params} can be
normalized in a universal way with a natural temperature scale in terms of $T_m$ and $T_{LS}$, and a natural time scale adequate for the system 
(for a fixed system size).

\section*{Acknowledgements}
This work is supported by FONDECYT grant 1140514 and FONDECYT Iniciación grant 11150279. CL also acknowledges support
from Proyecto Inserción PAI-79140025 and UNAB DI-1350-16/R. JP also acknowledges partial support from FONDECYT Iniciación
11130501 and UNAB DI-15-17/RG.

\appendix

\section{Bézier model for the gamma parameters}
\label{appendix_bezier}

We described the functions $k(T)$ and $\theta(T)$ by expanding them in a polynomial basis known as Bernstein polynomials.
This expresses the curves $k$ and $\theta$ as Bézier curves, parameterized by

\begin{eqnarray}
k(T) = \sum_{i=0}^{24} A_i\binom{n}{i}\mathcal{T}^i(1-\mathcal{T})^{n-i}, \nonumber \\
\theta(T) = \sum_{i=0}^{24} B_i\binom{n}{i}\mathcal{T}^i(1-\mathcal{T})^{n-i},
\label{eq_bezier}
\end{eqnarray}
where $\mathcal{T}$ is a normalized temperature (between 0 and 1), defined for convenience as

\begin{equation}
\mathcal{T} = \frac{T - T_\text{min}}{T_\text{max}-T_\text{min}}.
\end{equation}

In our case we have used $T_{\min}$=6200 K and $T_{\max}$=6700 K. For this model the hyperparameters describing $k$ and
$\theta(T)$ are the coefficients $A_i$ and $B_i$, with $i$=0,\ldots,24, leading to a total of 50 parameters.

\section{Markov Chain Monte Carlo}
\label{appendix_MC}

We employed a log-likelihood function divided into two terms. The first involves the data from melting times at a temperature 
$T_i$, and is given by

\begin{eqnarray}
\ln P(M_i|k, \theta) = n_i\Big(-\frac{t_i}{\theta(T_i)}+(k(T_i)-1)L_i \nonumber \\ 
-\ln \Gamma_{\text{inc}}(k(T_i); 0 \rightarrow \frac{\tau_0}{\theta(T_i)}) -k(T_i)\ln \theta(T_i)\Big),
\end{eqnarray}
where $n_i$ is the number of ocurrences of melting at $T_i$ (from a total of $N_i$), $t_i$ and $L_i$ denote the sample
averages at $T_i$ of $t_w$ and $\ln t_w$, respectively. The second contribution to the total log-likelihood includes the
number of realizations of melting $n_i$ and the total number of realizations $N_i$ at temperature $T_i$, and is given by

\begin{eqnarray}
\ln P(F_i|k, \theta) = n_i\ln p_i - (N_i-n_i)\ln (1-p_i) \nonumber \\
+ \ln \left[\frac{N_i!}{n_i!(N_i-n_i)!}\right],
\end{eqnarray}
with $p_i=P(t_w < \tau_0|T_i)$ the predicted probability of melting at temperature $T_i$. The logarithm of the posterior 
distribution of parameter functions $k$ and $\theta$ is then given by 

\begin{equation}
\ln P(k, \theta|\{M_i, F_i\}) = \sum_{i=1}^m\Big(\ln P(M_i|k, \theta) + \ln P(F_i|k, \theta)\Big),
\end{equation}
with flat uninformative priors. This logarithm of the posterior distribution was sampled using a Markov Chain Monte Carlo 
(MCMC) methodology, namely Metropolis-Hastings sampling~\cite{Gamerman2006}, over the hyperparameters present in the
models given by Eqs. \ref{eq_model3}, \ref{eq_model4} and \ref{eq_bezier}. 


\end{document}